\documentclass[aps,prd,onecolumn,showpacs,groupedaddress,nofootinbib]{revtex4-2}
\usepackage{graphicx}
\usepackage{dcolumn}
\usepackage{bm}
\usepackage{color}
\usepackage{longtable}
\usepackage{supertabular}
\usepackage[T1]{fontenc}
\usepackage{epstopdf}
\usepackage{amsmath}
\usepackage{amssymb}
\usepackage{setspace}
\usepackage{multirow}
\usepackage{booktabs}
\usepackage[section]{placeins}
\usepackage{mathrsfs}
\usepackage{hyperref}
\usepackage{adjustbox}

\makeatletter

\newcommand{\Rmnum}[1]{\expandafter\@slowromancap\romannumeral #1@} 
\newcommand{\bq}{\begin{equation}}
\newcommand{\eq}{\end{equation}}
\newcommand{\bqn}{\begin{eqnarray}}
\newcommand{\eqn}{\end{eqnarray}}
\newcommand{\nb}{\nonumber}

\makeatother

\begin{document}

\title{On the instability of the fundamental mode of the Regge-Wheeler effective potential}

\author{Shui-Fa Shen$^{1,2}$}
\author{Guan-Ru Li$^{3}$}
\author{Ramin G. Daghigh$^{4}$}
\author{Jodin C. Morey$^{5}$}
\author{Michael D. Green$^{4}$}
\author{Qiyuan Pan$^{6}$}
\author{Cheng-Gang Shao$^{7}$}
\author{Wei-Liang Qian$^{8,3,9}$}\email[E-mail: ]{wlqian@usp.br}

\affiliation{$^{1}$School of Intelligent Manufacturing, Zhejiang Guangsha Vocational and Technical University of Construction, 322100, Jinhua, Zhejiang, China}
\affiliation{$^{2}$Hefei Institutes of Physical Science, Chinese Academy of Sciences, 230031, Hefei, Anhui, China}
\affiliation{$^{3}$Faculdade de Engenharia de Guaratinguet\'a, Universidade Estadual Paulista, 12516-410, Guaratinguet\'a, SP, Brazil}
\affiliation{$^{4}$Metropolitan State University, Saint Paul, Minnesota, 55106, USA}
\affiliation{$^{5}$Le Moyne College, Syracuse, New York, 13214, USA}
\affiliation{$^{6}$Key Laboratory of Low Dimensional Quantum Structures and Quantum Control of Ministry of Education, Synergetic Innovation Center for Quantum Effects and Applications, and Department of Physics, Hunan Normal University, Changsha, Hunan 410081, China}
\affiliation{$^{7}$MOE Key Laboratory of Fundamental Physical Quantities Measurement, Hubei Key Laboratory of Gravitation and Quantum Physics, PGMF, and School of Physics, Huazhong University of Science and Technology, Wuhan 430074, China}
\affiliation{$^{8}$Escola de Engenharia de Lorena, Universidade de S\~ao Paulo, 12602-810, Lorena, SP, Brazil}
\affiliation{$^{9}$Center for Gravitation and Cosmology, School of Physical Science and Technology, Yangzhou University, Yangzhou 225009, China}

\begin{abstract}
It was recently pointed out that the fundamental mode of the black hole quasinormal spectrum might be subject to instability.
In particular, Cheung {\it et al.} showed that the fundamental mode of the Regge-Wheeler effective potential is unstable against an insignificant Gaussian metric perturbation, which, in turn, might substantially challenge the black hole spectroscopy.
This intriguing result has been interpreted by some authors as arising from essentially replacing the black hole's effective potential and its perturbation with two disjoint potential barriers.
We argue that such an analysis may have oversimplified the real physical scenario.
To be more precise, a metric perturbation planted farther away from the black hole horizon might not always be appropriately approximated by a disjoint minor barrier.
Particularly, for the perturbed P\"oschl-Teller potential, joint and disjoint metric perturbations might lead to drastically different stability properties for the low-lying modes.
Following this line of thought, this study conducts a refined analysis of the stability of the fundamental mode of the Regge-Wheeler effective potential by closely examining a few physically relevant ingredients.
Among other aspects, we retain the inverse-power-law tail of the Regge-Wheeler potential and consider a scenario in which the perturbation magnitude decreases with the radial coordinate to maintain a constant physical impact. 
While our analysis qualitatively confirms the main findings of previous studies, as the stability of the fundamental mode is primarily determined by the imaginary part of the quasinormal frequency, we show that specific features of both the effective potential at spatial infinity and the metric perturbation can have a sizable impact on the instability. 
In contrast, the spiral period, governed by the real part of the quasinormal frequency, appears largely insensitive to the details of the black hole metric or its perturbations. 
The analytic estimates are in reasonable agreement with the numerical results.
\end{abstract}

\date{Jan. 5th, 2026}

\maketitle

\newpage
\section{Introduction}\label{sec1}

The concept of black hole spectral instability was first pointed out by Nollert and Price~\cite{agr-qnm-instability-02, agr-qnm-instability-03} and Aguirregabiria and Vishveshwara~\cite{agr-qnm-27, agr-qnm-30}.
It was shown that insignificant modifications to the black hole metric, such as an approximation of the Regge-Wheeler effective potential in terms of a series of step functions, will drastically deform the quasinormal mode~\cite{agr-qnm-review-01, agr-qnm-review-02, agr-qnm-review-03} (QNM) spectrum.
Unlike the stability of normal modes for self-adjoint systems, it was demonstrated that the black hole QNM spectrum can be easily deformed due to the non-Hermitian nature of the underlying dissipative gravitational systems. 
This finding challenges the intuition that an insignificant modification to the effective potential is unlikely to have a significant impact on the resulting QNMs.

Further studies~\cite{agr-qnm-instability-11, agr-qnm-lq-03} suggested that even in the presence of a moderate discontinuity, the asymptotic behavior of the QNM spectrum remains significantly modified. 
Specifically, high overtones were found to lie almost parallel to the real frequency axis rather than ascending the imaginary frequency axis as for most black hole metrics~\cite{agr-qnm-continued-fraction-12, agr-qnm-continued-fraction-23}. 
As QNMs can be attributed to the poles of Green's function~\cite{agr-qnm-21, agr-qnm-29}, the largely uniform frequency-domain distribution of these poles implies~\cite{agr-qnm-echoes-20, agr-strong-lensing-correlator-15, agr-qnm-echoes-45, agr-qnm-instability-65} that this phenomenon is closely associated with the black hole echoes in the time-domain, a feature that occurs in the late-stage waveform first speculated by Cardoso {\it et al.}~\cite{agr-qnm-echoes-01, agr-qnm-echoes-review-01}.
Moreover, to a certain degree, the spectral instability undermines the general understanding that the unique characteristic of QNMs is primarily determined by the spacetime properties surrounding the black hole, which subsequently carries unambiguous information about the spacetime geometry near the event horizon.
Conversely, metric perturbations that trigger spectral instability can be placed far from the compact object.
Specifically, black hole spectral instability is observed to persist regardless of the discontinuity's distance from the horizon or its magnitude.
In the literature, these types of metric perturbations are dubbed as {\it ultraviolet}~\cite{agr-qnm-instability-07} due to their small scale or high frequency nature. 
Further expanding on these results, Jaramillo {\it et al.}~\cite{agr-qnm-instability-07, agr-qnm-instability-13} systematically explored the implications of spectral instability by analyzing the effects on the pseudospectrum of random and sinusoidal perturbations applied to the effective potential.
Their analyses revealed that the boundary of the pseudospectrum moves closer to the real frequency axis, thus forming a general picture of a universal instability of high-overtone modes triggered by ultraviolet perturbations.

The significance of spectral instability lies in its observational implications, particularly in the context of black hole spectroscopy~\cite{agr-BH-spectroscopy-review-04}.
This subject is now of practical importance given the successful detections of gravitational waves emanating from the binary mergers accomplished by the ground-based LIGO and Virgo collaboration~\cite{agr-LIGO-01, agr-LIGO-02, agr-LIGO-03, agr-LIGO-04}, as well as the ongoing space-borne projects, such as LISA~\cite{agr-LISA-01}, TianQin~\cite{agr-TianQin-01, agr-TianQin-Taiji-review-01}, and Taiji~\cite{agr-Taiji-01}.
In particular, the feasibility of testing the no-hair theorem by direct observation of ringdown waveforms has been studied~\cite{agr-TianQin-05}.
In realistic astrophysical circumstances, the source of gravitational waves is hardly isolated.
Therefore, for ``dirty'' black holes~\cite{agr-bh-thermodynamics-12, agr-qnm-33, agr-qnm-34, agr-BH-spectroscopy-10}, spectral instability poses a potential challenge for the gravitational data inference procedure.

In the literature, the interrelated topics of spectral instability, echoes, and causality have attracted the interest of many authors~\cite{agr-qnm-instability-08, agr-qnm-instability-13, agr-qnm-instability-14, agr-qnm-instability-15, agr-qnm-instability-16, agr-qnm-instability-18, agr-qnm-instability-19, agr-qnm-instability-26, agr-qnm-echoes-22, agr-qnm-echoes-29, agr-qnm-echoes-30, agr-qnm-instability-23, agr-qnm-instability-29, agr-qnm-instability-32, agr-qnm-instability-33, agr-qnm-instability-43, agr-qnm-echoes-35} (see~\cite{spectral-instability-review-20} for a brief review of recent developments).
Notably, Cheung {\it et al.}~\cite{agr-qnm-instability-15} pointed out that even the fundamental mode can be destabilized under rather generic perturbations.
By introducing a small perturbation to the Regge-Wheeler effective potential, it was shown~\cite{agr-qnm-instability-15} that, 
as the magnitude of the perturbation increases,
the fundamental mode undergoes an outward spiral.
In Refs.~\cite{agr-qnm-instability-32, agr-qnm-instability-56, agr-qnm-instability-58}, the numerical findings on the instability of the fundamental mode were interpreted in terms of the instability of an approximated case where the metric perturbation presents itself as a minor disjoint potential barrier, a simplified scenario which was initially proposed in the original study~\cite{agr-qnm-instability-15}. 
Specifically, for two disjoint effective potentials, the overall transmission coefficients can be formally written in terms of the coefficients of the individual potential barriers.
Under the approximation that the perturbative potential barrier is small in size, the deviations $\delta\omega_n$ in frequency from the original quasinormal modes satisfies~\cite{agr-qnm-instability-32, agr-qnm-instability-56, agr-qnm-instability-58, agr-qnm-instability-55}
\bqn
\delta\omega_n^\mathrm{disj}\equiv \omega-\omega_n =\mathcal{J}_n e^{-2\mathrm{Im}\omega_n x_c} e^{2i \mathrm{Re}\omega_n x_c} ,\label{DoubleSqureBarrier}
\eqn
where $\mathrm{Im}\omega_n$ and $\mathrm{Re}\omega_n$ are the imaginary and real parts of the original quasinormal frequency $\omega_n$ with overtone $n$, $\omega$ is the perturbed quasinormal mode corresponding to $\omega_n$, $x_c$ measures the distance between the two potential barriers, and the specific form of $\mathcal{J}_n$ is governed by the background metric and its perturbation, which vanish as the magnitude of the perturbation goes to zero.
As long as $\mathcal{J}_n$ is a moderate function of $x_c$, the behavior of $\delta\omega_n$ is mostly determined by the exponential factor $e^{-2\mathrm{Im}\omega_n x_c} e^{2i \mathrm{Re}\omega_n x_c}$.
In particular, $e^{-2\mathrm{Im}\omega_n x_c}$ primarily governs the magnitude of the deviation $\delta\omega_n$, while $e^{2i \mathrm{Re}\omega_n x_c}$ controls its phase.
Since $\mathrm{Im}\omega_n <0$, the magnitude of the deviation increases exponentially, giving rise to the instability.
Also, the phase $2\mathrm{Re}\omega_n x_c$ is linear in $x_c$, and therefore the spiral rotates in the counterclockwise direction when  $\mathrm{Re}\omega_n >0$.
When $|\mathrm{Im}\omega_n| < |\mathrm{Re}\omega_n|$, the rotation is relatively faster, therefore the spiral is visually more pronounced for the low-lying modes.
On the other hand, for high overtones $n\gg 1$, since $|\mathrm{Im}\omega_n|$ becomes more significant, the deviation is more sizable.
The above observations have been shown to be in quantitative agreement with numerical results~\cite{agr-qnm-instability-55}.
It is worth pointing out that once $\delta\omega_n$ becomes too large, the approximations leading to Eq.~\eqref{DoubleSqureBarrier} will break down and invalidate the observed spiral.

However, the perturbation introduced in~\cite{agr-qnm-instability-15} was placed in a continuous fashion in the Regge-Wheeler effective potential.
An analytic description of the QNMs for such a perturbation is not straightforward for the Regge-Wheeler potential.  Instead, these perturbations are analyzed using the Pöschl-Teller effective potential, where reasonable analytic approximations can be made.
It can be shown that the deviation of quasinormal frequencies reads~\cite{agr-qnm-instability-55}:
\bqn
\delta\omega_n^{\mathrm{PT}} = \mathcal{J}_n e^{(2i\omega_n-2\kappa) x_c}
=\mathcal{J}_n e^{\left(2n-1\right)\kappa x_c} e^{2 i \sqrt{V_0-\frac{\kappa^2}{4}} x_c} ,\label{fundamentalInsPT}
\eqn
where
$x_c$ is the coordinate where a minor step or a truncation is introduced to the unperturbed effective potential, $\kappa$ and $V_0$ are two positive parameters of the (inverse) Pöschl-Teller effective potential
\begin{eqnarray}
{V}_\mathrm{PT}=\frac{V_0}{\cosh ^2(\kappa x)} ,\label{V_PT}
\end{eqnarray}
and again, $\mathcal{J}_n$ is determined by the background metric whose specific form is not relevant for the physical outcome.
For this particular example, one notices that as long as $n > \frac12$, the spiral will occur and most conclusions drawn previously from Eq.~\eqref{DoubleSqureBarrier} remain unchanged.
Surprisingly, the fundamental mode $(n=0)$ is found to be stable against the  perturbation. 
Also, it is worth pointing out that, unlike Eq.~\eqref{DoubleSqureBarrier} where the QNMs are attributed to the zero of the Wronskian (or the transmission amplitude), the QNMs in this case are owing to the poles in the reflection amplitude.
For a stable fundamental mode, the counterclockwise spiral persists and has been verified quantitatively by numerical calculations.

Now, one might argue that the difference in stability for the fundamental mode in the P\"oschl-Teller and Regge-Wheeler potentials is due to the properties of the spacetime curve in asymptotic spatial infinity, where the perturbation is placed.
However, upon closer observation, the magnitude of the Regge-Wheeler effective potential decreases with an increasing radial coordinate according to an inverse power law.
The Pöschl-Teller effective potential, on the other hand, is suppressed more significantly, following an exponential form.
Following this line of thought, a disjoint perturbation corresponds to an even more exaggerated suppression of the potential.
In this regard, it is not straightforward to justify the instability of the fundamental mode of the Regge-Wheeler effective potential by comparing it to two disjoint potential barriers, as is done in \cite{agr-qnm-instability-15}.

A related question arises regarding the physical meaning of the perturbative bump's magnitude.
In~\cite{agr-qnm-instability-59}, it was pointed out that the strength of the deformation in the effective potential should be measured by its energy norm, not merely its peak amplitude.
Physically, what matters is the total energy contained within this bump.
The perturbation's impact on QNMs can be captured mathematically through the notion of $\epsilon$-pseudospectrum, which is an open set that collects all the QNMs for metric deformations up to a certain strength $\epsilon$, computed via the energy scalar product~\cite{agr-qnm-instability-08}.
Numerical results indicate that, with fixed amplitude and parameters, this energy norm grows proportionally to $r^2$.
Therefore, the apparent progressive destabilization of low-lying modes as the perturbative bump shifts farther from the compact object stems partly from this rapid energy norm increase under constant height of the perturbative bump.
Specifically, recent calculations with the modified P\"oschl-Teller potential~\cite{agr-qnm-instability-84} reveal that such destabilization diminishes significantly when accounting for this effect.

The present study is motivated by the above considerations.
We elaborate on the stability of the fundamental mode by a refined approximation of the Regge-Wheeler effective potential and the perturbation.
In particular, we devise a  perturbation that is physically appropriate. 
Our analysis consists of the following key ingredients:
\begin{itemize}
\item The approximate form of the Regge-Wheeler effective potential must reproduce the main features of the QNMs, particularly the fundamental mode, which should closely match its original counterpart.
\item The wavefunction must be largely accessible analytically, with the homogeneous version of the master equation admitting exact closed-form solutions.
\item The perturbation must be physically relevant, with a non-constant magnitude.
In particular, the deformation's magnitude decreases with the radial coordinate to maintain constant physical impact.
\item The model must provide an analytic description of the fundamental mode's deviation as the perturbation's location shifts away from the black hole.
\end{itemize}

The remainder of the paper is organized as follows.
In the next section, we briefly lay out the theoretical framework for calculating small deviations of QNMs and elaborate on an approximate form of the Regge-Wheeler effective potential and its perturbation.
In Sec.~\ref {sec3}, by employing the analytic closed-form solutions of the homogeneous equation, we analyze the stability of the fundamental mode by a semi-analytic approach.
Numerical calculations supporting our theoretical findings are presented in Sec.~\ref{sec4}.
The last section includes further discussions and concluding remarks.

\section{Quasinormal modes and an approximation for the Regge-Wheeler potential}\label{sec2}

We first present the theoretical setup of our present study.
For the Schwarzschild black hole, the study of black hole perturbations can be simplified by exploring the master equation~\cite{agr-qnm-review-03,agr-qnm-review-06},
\begin{eqnarray}
\frac{\partial^2}{\partial t^2}\Psi(t, x)+\left(-\frac{\partial^2}{\partial x^2}+V_\mathrm{RW}\right)\Psi(t, x)=0 ,
\label{master_eq_ns}
\end{eqnarray}
where $x$ is the tortoise coordinate, which is related to the radial coordinate $r$ according to $dx = dr/f(r)$ where $f(r)$ is the metric function given by
\bqn
f=1-r_h/r ,
\label{f_master}
\eqn
with the horizon radius $r_h=2M$ for a black hole of mass $M$.
The Regge-Wheeler potential $V_\mathrm{RW}$ is
\bqn
V_\mathrm{RW}=f\left[\frac{\ell(\ell+1)}{r^2}+(1-{s}^2)\frac{r_h}{r^3}\right],
\label{V_RW}
\eqn
where ${s}$ is the spin and  $\ell$ is the angular momentum of the perturbation.

In the frequency domain, Eq.~\eqref{master_eq_ns} has the form~\cite{agr-qnm-review-02},
\begin{eqnarray}
\left[-\omega^2-\frac{d^2}{dx^2}+V_\mathrm{RW}\right]\widetilde{\Psi}(\omega, x)=0.
\label{pt_homo_eq}
\end{eqnarray}
The QNM frequencies can be obtained by evaluating the zeros of the Wronskian
\begin{eqnarray}
W(\omega)\equiv W(g,h)=g(\omega,x)h'(\omega, x)-h(\omega,x)g'(\omega,x) ,
\label{pt_Wronskian}
\end{eqnarray}
where $'\equiv d/dx$, and $h$ and $g$ are the solutions to Eq.~\eqref{pt_homo_eq} 
with appropriate boundary conditions, namely,
\begin{eqnarray}
\begin{array}{cc}
h(\omega, x)\sim e^{-i\omega x}    &  x\to -\infty  \cr\\
g(\omega, x)\sim e^{i\omega x}     &  x\to +\infty  
\end{array} .
\label{pt_boundary}
\end{eqnarray}

A feasible strategy for evaluating small deviations of QNMs was presented in~\cite{agr-qnm-instability-55}, which is summarized as follows.
The Wronskian Eq.~\eqref{pt_Wronskian} remains zero in the presence of the metric perturbation, which is influenced by two factors.  
First, as long as the deviation $\delta\omega_n$ is small, one expects that the deviations of most quantities that constitute the Wronskian can be reasonably captured by the first term of a Taylor expansion in $\omega$, and therefore are linear in $\delta\omega_n$. 
Secondly, the modification to Eq.~\eqref{pt_Wronskian} also involves $x_c$, the location of the perturbation.
Moreover, the latter typically enters the expressions in the form of an exponential function $e^{2i x_c\omega}$, which can be attributed to the spatial transformation operation involved. 
For the fundamental mode, one typically has $x_c\gg |\omega|$, and the term $e^{2i x_c\omega}$ is primarily governed by the variation of $x_c$ (not $\omega$).
As a result, the QNMs are obtained when the above two sources of modification to the Wronskian reach a balance so that it remains zero.
In practice, however, such an approach only becomes viable when one possesses analytic closed (at least, asymptotic) forms for the solution of the homogeneous Eq.~\eqref{pt_homo_eq}.

In light of the above discussions, we aim to devise a reasonable approximation to the Regge-Wheeler effective potential $V_\mathrm{RW}$.
Following~\cite{agr-qnm-instability-11}, we consider the approximation 
\bqn
V_\mathrm{Apx} = \left\{ \begin{array}{ll}
	 \left[\ell(\ell+1)+(1-s^2)\right]e^{x-1}  & \mbox{$x < x_0$}~\\  \\
  V_0 & \mbox{$x_0 \leq x < x_1$}~\\  \\
	 {\ell(\ell+1)}/{x^2}  & \mbox{$x \geq x_1$,}
\end{array}
\right.        
\label{V_RW_app0}
\eqn
where $V_0$ is taken to be the maximum of the Regge-Wheeler potential, forming a plateau between the interval $[x_0, x_1]$.
It is readily verified that at the spatial boundary $x\to \pm\infty$, Eq.~\eqref{V_RW_app0} asymptotically approaches Eq.~\eqref{V_RW} with $r_h=1$.  This potential approximates the Regge-Wheeler potential for a Schwarzschild black hole of $r_h=1$ with the following parameters $\ell=2$, $s=0$, $x_0=-0.9575$, and $x_1=2.4637$.  
Hereafter, all quantities are expressed in geometric units where $c=G=1$ and $r_h=1$.

\begin{figure}[h]
\begin{tabular}{cc}
\begin{minipage}{250pt}
\centerline{\includegraphics[width=1.0\textwidth]{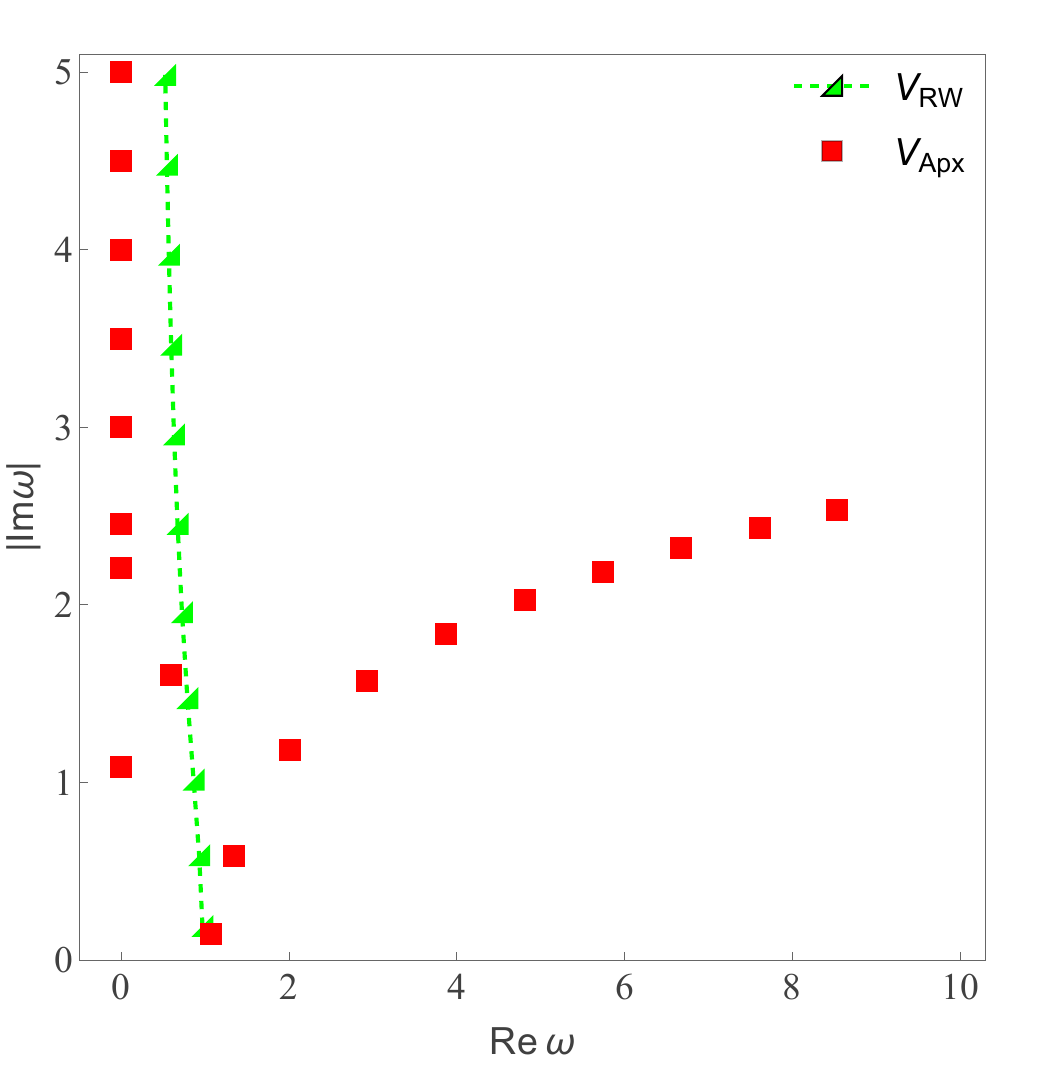}}
\end{minipage}
&
\begin{minipage}{250pt}
\centerline{\includegraphics[width=1.0\textwidth]{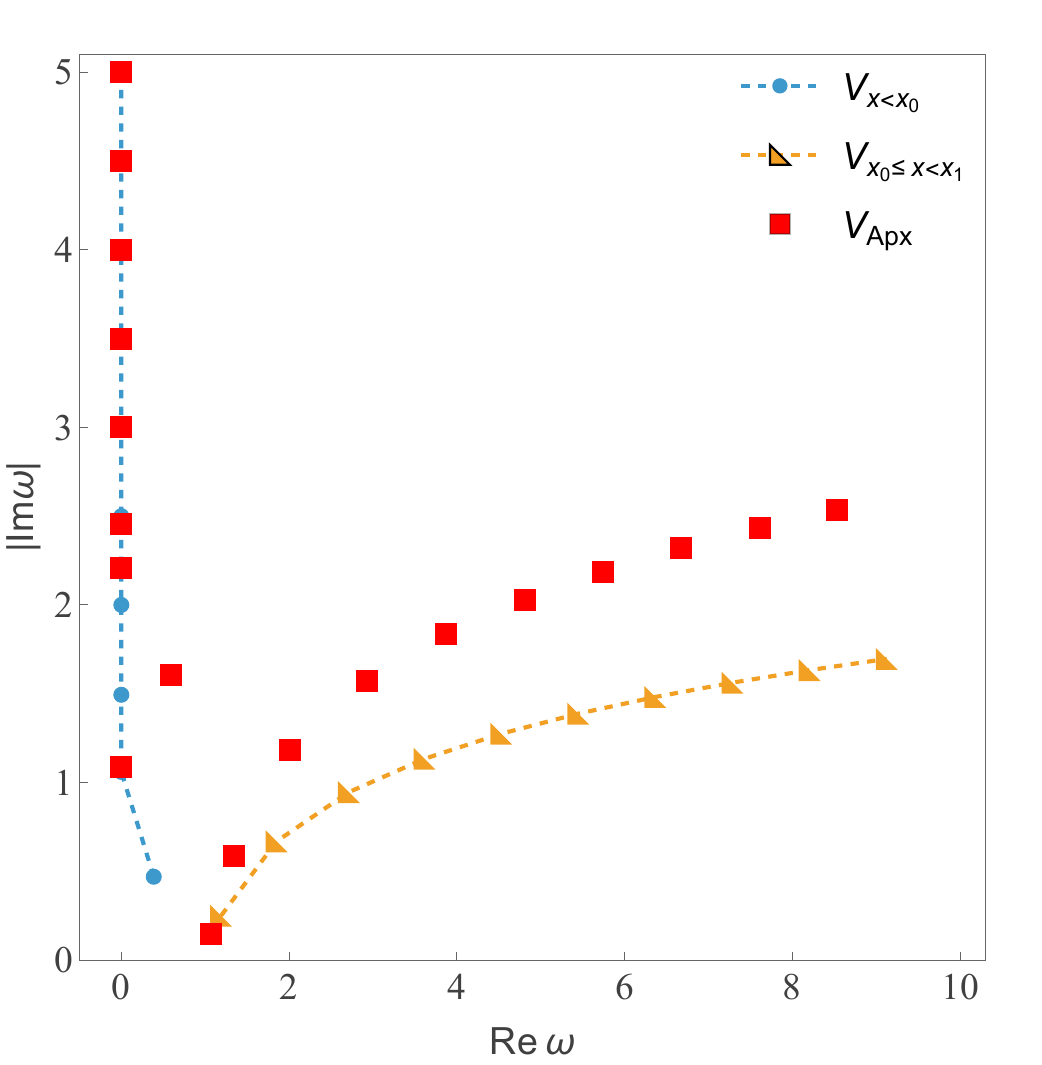}}
\end{minipage}
\end{tabular}
\renewcommand{\figurename}{Fig.}
\caption{The resulting QNMs of the approximate Regge-Wheeler effective potential compared against those of the original Regge-Wheeler potential.
Left: The calculated QNMs of the approximate effective potential Eq.~\eqref{V_RW_app0} (shown in filled red squares) compared with those of the Regge-Wheeler potential Eq.~\eqref{V_RW} (indicated by green triangles).
Right: The calculated QNMs of the approximate effective potential (shown in filled red squares) compared with those obtained by separated regions of the piecewise function defined in the first and second lines of Eq.~\eqref{V_RW_app0} (indicated by filled blue circles and orange triangles).
The numerical calculations are carried out using the wavefunctions given by Eqs.~\eqref{RWApxf}-\eqref{RWApxg}, where one assumes the parameters $\ell=2$, $s=0$, $x_0=-0.9575$, and $x_1=2.4637$.
One notes that the fundamental mode of the approximate potential is very close to that of the original black hole.
}
\label{RW_app_QNMs} 
\end{figure}

As shown in the left panel of Fig.~\ref{RW_app_QNMs}, the QNMs of the approximate potential split into two branches. 
The fundamental mode sits very close to that of the original Schwarzschild black hole.
For high overtones, one branch has the feature $|\mathrm{Im}\omega_n| \gg \mathrm{Re}\omega_n$, displaying a similar trend as the Regge-Wheeler effective potential.
Comparing with the QNMs indicated by the blue solid circles shown in the right panel of Fig.~\ref{RW_app_QNMs}, it can be inferred that this branch can be primarily attributed to the asymptotic behaviour of the part of the effective potential near the horizon, which is consistent with the understanding that QNMs can be used to probe the properties of the spacetime near the horizon.
On the other hand there is a second branch of QNMs, which is not observed for the original black hole metric.
The emergence of such a branch can be attributed to the presence of a  discontinuity~\cite{agr-qnm-lq-03}, and can be qualitatively understood when comparing with the QNMs of a square barrier shown in the right panel by filled orange triangles.
Specifically, the plateau of length $L=x_1-x_0$ leads to a series of ``echo modes,'' and the distance along the real frequency axis between successive modes is given by $\Delta\omega=\pi/L\sim 1$, borrowing the derivation given in~\cite{agr-qnm-echoes-20, agr-qnm-instability-65}.
This feature is readily demonstrated in either the filled orange triangles or the branch of filled red squares just above the orange triangles.

An apparent advantage of the approximation Eq.~\eqref{V_RW_app0} is that the wavefunctions are analytic.
Specifically, the wavefunctions satisfying the boundary conditions Eq.~\eqref{pt_boundary} possess the forms~\cite{agr-qnm-instability-11, agr-qnm-tail-06}
\bqn
\widetilde{\Psi}_1(\omega,x) =	D  (-1)^{-i\omega}\Gamma(1-2i\omega)I_{-2i\omega}\left(2\sqrt{\left[\ell(\ell+1)+(1-s^2)\right] e^{x-1}}\right) \ \ \ \ \mbox{$x < x_0$} ,
\label{RWApxf}
\eqn
\bqn
\widetilde{\Psi}_2(\omega,x) = E e^{i\sqrt{\omega^2-V_0}~x} + F e^{-i\sqrt{\omega^2-V_0}~ x} \ \ \ \ \mbox{$x_0 \le x < x_1$} ,
\eqn
and
\bqn
\widetilde{\Psi}_3(\omega,x) = G e^{i\frac{\pi}{2} (\ell-1)}\sqrt{\frac{\pi\omega x}{2}}H_{\ell+\frac12}^{(1)}(\omega x)\ \ \ \  \mbox{$x \ge x_1$} ,
\label{RWApxg}
\eqn
where $\omega$ is the complex QNM frequency to be determined, $I_{\nu}(z)$ is the modified Bessel function of the first kind and $H_\nu^{(1)}(z)$ denotes the Hankel function of the first kind.
The coefficients $D$, $E$, $F$, and $G$ are constants.
We note that Eqs.~\eqref{RWApxf} and~\eqref{RWApxg} satisfy the proper boundary conditions Eq.~\eqref{pt_boundary}.
The QNMs shown in Fig.~\ref{RW_app_QNMs} are obtained by numerically solving for the zeros of the Wronskian evaluated using Eqs.~\eqref{RWApxf}-\eqref{RWApxg} at the junction points $x_0$ and $x_1$.
One notes that the general solutions of the inverse-square-law potential in the Hankel functions can be simplified and rewritten using elementary functions when $\ell$ is an integer. 
Specifically, for $\ell=2$, we have
\bqn
\widetilde{\Psi}_3(\omega,x) = K e^{i\omega x}\left( -1 +\frac{3}{i\omega x} +\frac{3}{\omega^2 x^2}\right)\ \ \ \  \mbox{$x \ge x_1$} .
\label{RWApxg2}
\eqn
Furthermore, as elaborated in the next section, an analytic account of the modification to the fundamental mode due to metric perturbation is feasible by employing the strategy described above.

\section{A semi-analytic approach for the stability of perturbed Regge-Wheller potential}\label{sec3}

In this section, we introduce a perturbation to the approximate Regge-Wheeler effective potential Eq.~\eqref{V_RW_app0}, and analyze the impact on the fundamental mode.
For mathematical simplicity, we first consider the case where the effective potential is truncated at $x_c$
\bqn
V_\mathrm{Apx}^\mathrm{truc} = \left\{ \begin{array}{ll}
	 \left[\ell(\ell+1)+(1-s^2)\right]e^{x-1}  & \mbox{$x < x_0$}~\\  \\
  V_0 & \mbox{$x_0 \leq x < x_1$}~\\  \\
	 {\ell(\ell+1)}/{x^2}  & \mbox{$x_1 \leq x < x_c$~}\\ \\
  0  & \mbox{$x \geq x_c$~.}
\end{array}
\right.        
\label{V_RW_app1}
\eqn
Although such a scenario might not be physically pertinent, it is mathematically simple, and most of the obtained results can be straightforwardly generalized to a more relevant case:
\bqn
V_\mathrm{Apx}^\mathrm{step} = \left\{ \begin{array}{ll}
	 \left[\ell(\ell+1)+(1-s^2)\right]e^{x-1}  & \mbox{$x < x_0$}~\\  \\
  V_0 & \mbox{$x_0 \leq x < x_1$}~\\  \\
	 {\ell(\ell+1)}/{x^2}  & \mbox{$x_1 \leq x < x_c$~}\\ \\
  {\ell'(\ell'+1)}/{x^2}  & \mbox{$x \geq x_c$~}
\end{array}
\right.        
\label{V_RW_app2}
\eqn
The effective potential defined in Eq.~\eqref{V_RW_app2} can be viewed as a metric perturbation, where an infinitesimally thin mass shell is added at $x=x_c$.
Similar to~\cite{agr-qnm-instability-15, agr-qnm-instability-55}, we consider the case where the perturbation is placed farther away from the black hole, so that $x_c \gg x_0$, and we assume $\ell'\gtrsim\ell$ so the perturbation is not too large.  One enforces that the discontinuity at $x_c$ is proportional to $1/r_c^2\sim 1/x_c^2$, which corresponds to a scenario where the mass of the spherical shell remains constant as $x_c$ increases.

For Eqs.~\eqref{V_RW_app1} and~\eqref{V_RW_app2}, rather than having a disjoint potential, the perturbations are now placed on a continuous effective potential.
By using this approximation, we are able to avoid tedious calculations in the Regge-Wheeler case and to assess the problem analytically. 


We first analyze the migration of the fundamental modes of the approximated effective potential Eq.~\eqref{V_RW_app1}.
The wavefunctions in the four different regions possess the following forms
\bqn
\widetilde{\Psi}_1(\omega,x) =	D  (-1)^{-i\omega}\Gamma(1-2i\omega)I_{-2i\omega}\left(2\sqrt{\left[\ell(\ell+1)+(1-s^2)\right] e^{x-1}}\right) \ \ \ \ \mbox{$x < x_0$} ,
\label{RWAsolf}
\eqn
\bqn
\widetilde{\Psi}_2(\omega,x) = E e^{i\sqrt{\omega^2-V_0}~x} + F e^{-i\sqrt{\omega^2-V_0}~ x} \ \ \ \ \mbox{$x_0 \le x < x_1$} ,
\eqn
\bqn
\widetilde{\Psi}_3(\omega,x) = G e^{i\frac{\pi}{2} (\ell-1)}\sqrt{\frac{\pi\omega x}{2}}H_{\ell+\frac12}^{(1)}(\omega x)
+K e^{-i\frac{\pi}{2} (\ell-1)}\sqrt{\frac{\pi\omega x}{2}}H_{\ell+\frac12}^{(2)}(\omega x)\ \ \ \  \mbox{$x_1 \leq x < x_c$} ,
\label{RWAsoluH}
\eqn
and
\bqn
\widetilde{\Psi}_4(\omega,x) = N e^{i\omega x}  \ \ \ \  \mbox{$x \geq x_c$} ,
\label{RWAsolg}
\eqn
where the coefficients $D$, $E$, $F$, $G$, $K$, and $N$ are constants, and $H_\nu^{(2)}(z)$ denotes the Hankel function of the second kind.
It is verified that Eqs.~\eqref{RWAsolf} and~\eqref{RWAsolg} satisfy the proper boundary conditions Eq.~\eqref{pt_boundary}.
Again, for $\ell=2$, Eq.~\eqref{RWAsoluH} simplifies to
\bqn
\widetilde{\Psi}_3(\omega,x) = G e^{i\omega x}\left(-1 +\frac{3}{i\omega x} +\frac{3}{\omega^2 x^2}\right)
+K e^{-i\omega x}\left(-1 -\frac{3}{i\omega x} +\frac{3}{\omega^2 x^2}\right)\ \ \ \  \mbox{$x_1 \leq x < x_c$} ,
\label{RWAsolu}
\eqn
To evaluate the Wronskian at the point of discontinuity $x=x_0$, one uses the properties of the Bessel functions
\bqn
\frac{\partial I_\nu(z)}{\partial z} = \frac12\left(I_{\nu-1}(z)+I_{\nu+1}(z)\right) ,
\eqn
and
\bqn
I_\nu(z) \sim \frac{e^z}{\sqrt{2\pi z}} \left(1 - \frac{4 \nu^2 - 1}{8z} + \frac{\left(4 \nu^2 - 1\right) \left(4 \nu^2 - 9\right)}{2! (8z)^2} + \cdots \right) \ \ \ \ \text{for }\left|\arg z\right|<\frac{\pi}{2},
\eqn
and large $|z|$\footnote{When $|z|$ cannot be considered large, Eq.~\eqref{theFirstJunCon} must be replaced by its exact form.
Nonetheless, we argue that the main conclusion of this section remains valid, since the r.h.s.\ of Eq.~\eqref{theFirstJunCon} is independent of $x_c$, leaving the $x_c$ dependence unchanged in both the exponential and the prefactor $\mathcal{J}_n$ from Eqs.~\eqref{SpiralRWApp1} and~\eqref{SpiralRWApp2}.}.
Here, $z=2\sqrt{\left[\ell(\ell+1)+(1-s^2)\right] e^{x-1}}$, $\nu=-2i\omega$, and we have
\bqn
\frac{dz}{dx}=\frac12 z .
\eqn
One proceeds to evaluate the derivative
\bqn
\frac{I_\nu'}{I_\nu}
&=&\frac{dz}{dx}\frac{\frac12\left(I_{\nu-1}(z)+I_{\nu+1}(z)\right)}{I_\nu} \nb\\
&=&\frac{z}{2}\frac{\frac12\left[\frac{e^z}{\sqrt{2\pi z}}\left(1-\frac{4(\nu-1)^2-1}{8z}\right)+\frac{e^z}{\sqrt{2\pi z}}\left(1-\frac{4(\nu+1)^2-1}{8z}\right)\right]}{\frac{e^z}{\sqrt{2\pi z}}\left(1-\frac{4\nu^2-1}{8z}\right)} \nb\\
&=&\frac{z}{2}\frac{1-\frac{4\nu^2+3}{8z}}{1-\frac{4\nu^2-1}{8z}} 
\sim\frac{z}{2}\left(1-\frac{4\nu^2+3}{8z}+\frac{4\nu^2-1}{8z}\right)
=\frac{z}{2}\left(1-\frac{1}{2z}\right)
\eqn
which gives
\bqn
\frac{\widetilde{\Psi}_1'}{\widetilde{\Psi}_1}\simeq \frac{z}{2}\left(1-\frac{1}{2z}\right) .\label{theFirstJunCon}
\eqn
We also compute the ratios of the remaining wavefunctions
\bqn
\frac{\widetilde{\Psi}_2'}{\widetilde{\Psi}_2}=i\sqrt{\omega^2-V_0}\frac{E e^{i\sqrt{\omega^2-V_0}~x} - F e^{-i\sqrt{\omega^2-V_0}~ x}}{E e^{i\sqrt{\omega^2-V_0}~x} + F e^{-i\sqrt{\omega^2-V_0}~ x}} ,
\label{rhph}
\eqn
\bqn
\frac{\widetilde{\Psi}_3'}{\widetilde{\Psi}_3}=\frac{Ke^{-ix\omega}(6i-6x\omega-3ix^2\omega^2+x^3\omega^3)-Ge^{ix\omega}(-6i-6x\omega+3ix^2\omega^2+x^3\omega^3)}{x\left[Ke^{-ix\omega}(-3i+3x\omega+ix^2\omega^2)+iGe^{ix\omega}(-3+3ix\omega+x^2\omega^2)\right]}.
\label{rupu}
\eqn
and
\bqn
\frac{\widetilde{\Psi}_4'}{\widetilde{\Psi}_4}=i\omega .
\label{rgpg}
\eqn

The junction condition evaluated at $x=x_0$ between $\widetilde{\Psi}_1$ and $\widetilde{\Psi}_2$ yields
\bqn
\frac{E}{F} = e^{-2i x_0 \sqrt{\omega^2-V_0}} \frac{i\sqrt{\omega^2-V_0}+\frac{z_0}{2}\left(1-\frac{1}{2z_0}\right)}{i\sqrt{\omega^2-V_0}-\frac{z_0}{2}\left(1-\frac{1}{2z_0}\right)},
\label{rGH}
\eqn
where $z_0=2\sqrt{\left[\ell(\ell+1)+(1-s^2)\right] e^{x_0-1}}$.
Similarly, the junction condition evaluated at $x=x_c$ implies
\bqn
\frac{K}{G}=e^{2i x_c\omega} \frac{3(2i+x_c\omega)}{-6i+9x_c\omega+6ix_c^2\omega^2-2x_c^3\omega^3}
\sim  -e^{2i x_c\omega} \frac{3}{2\omega^2 x_c^2},
\label{rEF}
\eqn

The relation between Eqs.~\eqref{rGH} and~\eqref{rEF} can be established via the Wronskian through Eqs.~\eqref{rhph} and~\eqref{rupu} evaluated at $x=x_1$ which, after some algebra, is found to be
\bqn
\frac{K}{G} = \mathscr{F}\left(\frac{E}{F}, x_1, V_0, \omega\right)
\equiv e^{2ix_1\omega}\frac{6i+e^{2ix_1\sqrt{\omega^2-V_0}}\frac{E}{F}C_3+C_4}{-6i-e^{2ix_1\sqrt{\omega^2-V_0}}\frac{E}{F}C_1+C_2} ,
\label{rEFGH}
\eqn
where
\bqn
C_1 &=& 6i-3ix_1^2\omega\left(\omega+\sqrt{\omega^2-V_0}\right)+x_1^3\omega^2\left(\omega+\sqrt{\omega^2-V_0}\right)-3x_1\left(2\omega+\sqrt{\omega^2-V_0}\right) ,\nb\\
C_2 &=& -3x_1\left(-2\omega+\sqrt{\omega^2-V_0}\right)-3ix_1^2\omega\left(-\omega+\sqrt{\omega^2-V_0}\right)+x_1^3\omega^2\left(-\omega+\sqrt{\omega^2-V_0}\right) ,\nb\\
C_3 &=& 6i-3x_1\left(-2\omega+\sqrt{\omega^2-V_0}\right)+3ix_1^2\omega\left(-\omega+\sqrt{\omega^2-V_0}\right)+x_1^3\omega^2\left(-\omega+\sqrt{\omega^2-V_0}\right) ,\nb\\
C_4 &=& -3ix_1^2\omega \left(\omega+\sqrt{\omega^2-V_0}\right)-x_1^3\omega^2\left(\omega+\sqrt{\omega^2-V_0}\right)+3x_1\left(2\omega+\sqrt{\omega^2-V_0}\right) .\nb
\eqn
It is essential that $\mathscr{F}$ is not a function of the perturbation location $x_c$. 
As will become clear below, although the specific form of the function $\mathscr{F}$ is crucial for the QNM of the original black hole, it is irrelevant to its deformation. 
In particular, one finds that the equation governing the QNM is
\bqn
J(\omega, V_0, x_0, x_1, x_c) =e^{2i x_c\omega} ,
\label{eqDev}
\eqn
where
\bqn
J(\omega, V_0, x_0, x_1, x_c) =  \mathscr{F}\left(e^{-2i x_0\sqrt{\omega^2-V_0}} \frac{i\sqrt{\omega^2-V_0}+\frac{z_0}{2}\left(1-\frac{1}{2z_0}\right)}{i\sqrt{\omega^2-V_0}-\frac{z_0}{2}\left(1-\frac{1}{2z_0}\right)}, x_1, V_0, \omega\right) \mathscr{G}_0 (\omega, x_c),
\label{defJ}
\eqn
and
\bqn
\mathscr{G}_0 (\omega, x_c) = - \frac{2\omega^2 x_c^2}{3} .\label{DefScrG0}
\eqn

It is noted that the exponential form $e^{2ix_c\omega}$ evolves more drastically than any remaining part of the function, because $x_c\gg 1$ and $i\omega=-\mathrm{Im}\omega+i\mathrm{Re}\omega$ possess a positive real part since $\mathrm{Im}\omega <0$.
As discussed in the text, the solution of $\omega$ must be estimated from its deviation from the QNMs governed by the potential Eq.~\eqref{V_RW_app0}.
To fall back to the QNMs of the unperturbed metric, one replaces Eq.~\eqref{RWAsolu} with Eq.~\eqref{RWApxg2}
\bqn
\bar{g}(\omega,x) = E e^{i\omega x}\left(-1 +\frac{3}{i\omega x} +\frac{3}{\omega^2 x^2}\right)
\ \ \ \  \mbox{$x > x_1$} ,\nb
\eqn
and meanwhile casting away Eq.~\eqref{RWAsolg}.
By making use of the relation
\bqn
\frac{\bar{g}'}{\bar{g}}=\frac{6-6ix\omega-3x^2\omega^2+ix^3\omega^3}{x(-3+3ix\omega+x^2\omega^2)} ,
\label{rupgbar}
\eqn
it is not difficult to confirm that the QNMs $\omega_n$ of the effective potential are given by the roots (c.f. Eqs.~\eqref{rEFGH},~\eqref{eqDev} and~\eqref{defJ}) of
\bqn
\mathscr{F}\left(e^{-2i x_0 \sqrt{\omega^2-V_0}} \frac{i\sqrt{\omega^2-V_0}+\frac{z_0}{2}\left(1-\frac{1}{2z_0}\right)}{i\sqrt{\omega^2-V_0}-\frac{z_0}{2}\left(1-\frac{1}{2z_0}\right)}, x_1, V_0, \omega\right) = 0 ,
\eqn
which implies $J(\omega_n, V_0, x_0, x_1)=0$ and shall be subtracted from both sides of Eq.~\eqref{eqDev}.

Subsequently, following the strategy introduced in the previous section, small deviations of the modes as a function of $x_c$ are governed by
\begin{equation}
\delta\omega_n \sim\mathcal{J}_n e^{-2\mathrm{Im}\omega_n x_c} e^{2i\mathrm{Re}\omega_n x_c} ,\label{SpiralRWApp1}
\end{equation}
where one has
\begin{equation}
\mathcal{J}_n =  \left.\frac{\partial J}{\partial \omega}\right|_{\omega=\omega_n}^{-1} .
\end{equation}

We note that the above result implies a spiral in the low-lying modes very similar to that initially demonstrated in~\cite{agr-qnm-instability-15}, and analytically accounted for in the literature~\cite{agr-qnm-instability-32, agr-qnm-instability-56, agr-qnm-instability-58, agr-qnm-instability-55} and summarized above in Eq.~\eqref{DoubleSqureBarrier}.

Given the above derivation, we proceed to consider the more realistic case Eq.~\eqref{V_RW_app2}.
Now, since $\ell'\gtrsim\ell$ is not an integer, the corresponding wave function has to be written in terms of the Hankel function.
Specifically, in the place of Eq.~\eqref{RWAsolg}, we have
\bqn
g(\omega,x) = K e^{i\frac{\pi}{2} (\ell'-1)}\sqrt{\frac{\pi\omega x}{2}} H_{\ell'+\frac12}^{(1)}(\omega x)  \ \ \ \  \mbox{$x \geq x_c$} ,
\label{RWAsolg2}
\eqn
where $H_{\nu}^{(1)}(z)$ is the first type of Hankel function.
By using the asymptotic expansion of the Hankel function for large $|z|$, we can approximate
\bqn
g(\omega,x) \sim K e^{i\omega x} \left[-1+\frac{4\left(\ell'+\frac12\right)^2-1}{i8\omega x}+\frac{\left(4\left(\ell'+\frac12\right)^2-1\right)\left(4\left(\ell'+\frac12\right)^2-9\right)}{128\omega^2x^2}\cdots\right] 
\sim K e^{i\omega x} \left(-1-\frac{3i}{\omega x}+\frac{3}{\omega^2 x^2}-\frac{5i\epsilon}{2\omega x^2}\right) .\nb\\
\label{RWAsolg3}
\eqn
In deriving the last expression, recall the discussion after Eq.~\eqref{V_RW_app2} where the difference of $\ell$ and $\ell^\prime$ is interpreted as adding a thin shell of mass $\Delta M \ll M$  at $x=x_c$.  
In this case the relationship between $\Delta M$ and $\Delta \ell$ can be derived from Eq.~\eqref{V_RW} by replacing $M$ with $M+\Delta M$ giving 
\bqn
\frac{(\Delta \ell)^2}{r^2} \sim \frac{2\Delta M}{r^3} .
\eqn
If we set
\bqn
\Delta\ell \equiv \ell'-\ell = \frac{\epsilon}{x} ,\label{DeltaEll}
\eqn
where $\epsilon\sim \Delta M/(x\ell)$ is a small constant, and take $\ell=2$, we obtain the last expression in Eq.~\eqref{RWAsolg3}.
This gives
\bqn
\frac{g'}{g}\sim i\omega\left(1-\frac{3}{\omega^2x^2}-\frac{-3i+10\epsilon\omega}{\omega^3x^3}\right)  ,
\label{rgpg2}
\eqn
in place of Eq.~\eqref{rgpg}.

Applying the junction condition at $x=x_c$, this leads to the ratio
\bqn
\frac{F}{E}&=&e^{2i x_c\omega} \frac{9i+10\epsilon\omega(-3+3ix_c\omega+x_c^2\omega^2)}
{-(9i+30\epsilon\omega)
-12x_c^3\omega^3
-6ix_c^4\omega^4
+2x_c^5\omega^5
+6x_c\omega(3+5i\epsilon\omega)
-2x_c^2\omega^2(-9i+5\epsilon\omega)}\nb\\
&&\sim  e^{2i x_c\omega} \frac{9i+ 10 \epsilon\omega^3 x_c^2}{2\omega^5 x_c^5},
\label{rEF2}
\eqn
which is explicitly dependent on $\epsilon$.

Fortunately, since Eq.~\eqref{rEFGH} remains unchanged, one finally encounters a result that is formally identical to Eq.~\eqref{eqDev}
\bqn
\bar{J}(\omega, V_0, x_0, x_1, x_c) = e^{2i x_c\omega} ,
\label{eqDev2}
\eqn
where
\bqn
\bar{J}(\omega, V_0, x_0, x_1, x_c) =  \mathscr{F}\left(e^{-2i x_0\sqrt{\omega^2-V_0}} \frac{i\sqrt{\omega^2-V_0}+\frac{z_0}{2}\left(1-\frac{1}{2z_0}\right)}{i\sqrt{\omega^2-V_0}-\frac{z_0}{2}\left(1-\frac{1}{2z_0}\right)}, x_1, V_0, \omega\right)\mathscr{G}_1 (\omega, x_c) ,
\label{defJ2}
\eqn
and
\bqn
\mathscr{G}_1 (\omega, x_c) = \frac{2\omega^5 x_c^5}{9i+ 10 \epsilon\omega^3 x_c^2} .\label{DefScrG1}
\eqn

The difference between Eqs.~\eqref{DefScrG0} and~\eqref{DefScrG1} will not quantitatively affect the previous conclusion, since the deviation of the quasinormal frequency is primarily governed by the exponential factor $e^{2i x_c\omega}$.
All in all, one concludes that the instability of the fundamental mode remains unchanged as
\begin{equation}
\delta\omega_n \sim\bar{\mathcal{J}}_n e^{-2\mathrm{Im}\omega_n x_c} e^{2i\mathrm{Re}\omega_n x_c} ,\label{SpiralRWApp2}
\end{equation}
where one has
\begin{equation}
\bar{\mathcal{J}}_n =  \left.\frac{\partial \bar{J}}{\partial \omega}\right|_{\omega=\omega_n}^{-1} .
\end{equation}

Observing Eqs.~\eqref{defJ} and~\eqref{defJ2}, on the right-hand-sides, only the factors $\mathscr{G}_0$ and  $\mathscr{G}_1$ depend on $x_c$.
As they are rational functions for the two specific cases considered in the present study, their dependence on $x_c$ is not as sensitive as the exponential function $e^{2i x_c\omega}$.
Therefore, in deriving Eqs.~\eqref{SpiralRWApp1} and~\eqref{SpiralRWApp2}, one has essentially ignored the $x_c$ dependence of  $\mathscr{G}_0$ and  $\mathscr{G}_1$.
However, as shown in the next section, the main spiral behavior derived above can be substantially modified by considering the contribution from  $\mathscr{G}_0$ and  $\mathscr{G}_1$.
The latter, which essentially carries information on the spacetime metric at asymptotic spatial infinity and its perturbation, might play a crucial role in the spiral of the fundamental mode.

\section{Numerical results}\label{sec4}

\begin{figure}[h]
\begin{tabular}{cc}
\begin{minipage}{250pt}
\centerline{\includegraphics[height=0.9\textwidth, width=1.0\textwidth]{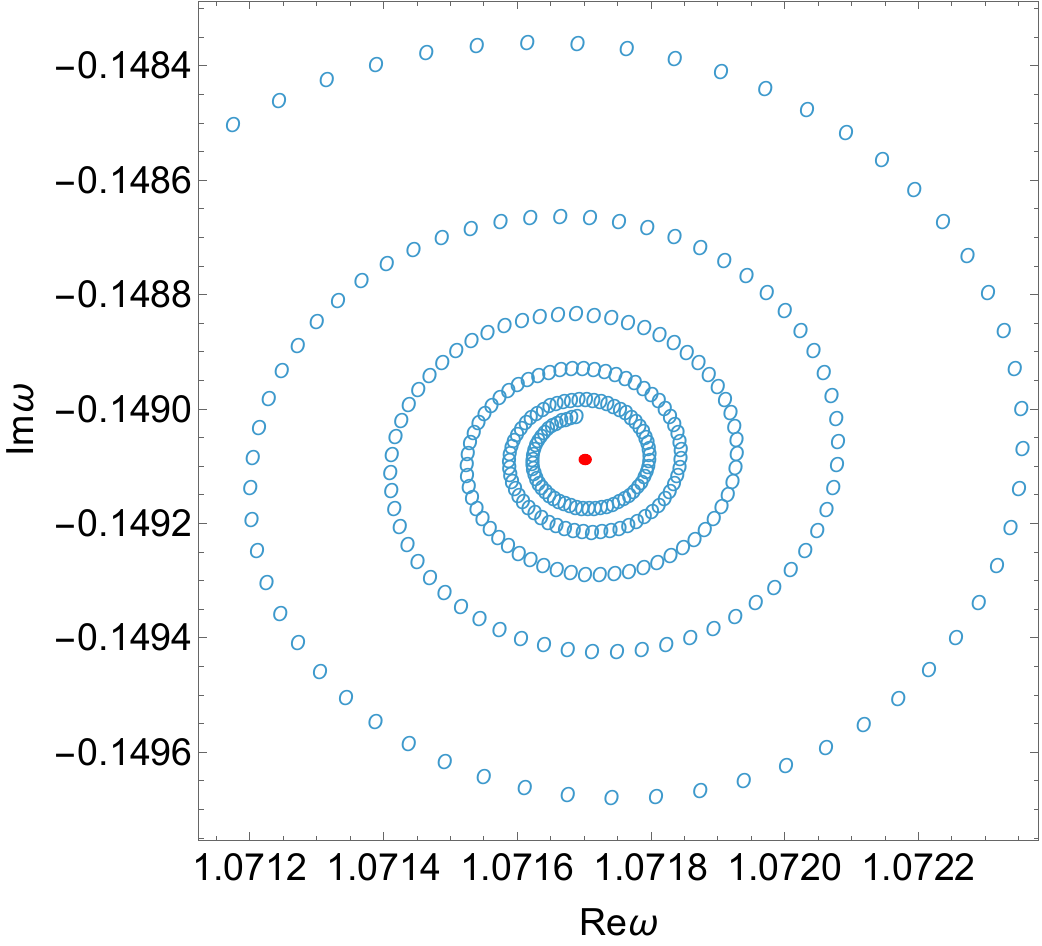}}
\end{minipage}
&
\begin{minipage}{250pt}
\vspace{-0.75cm}
\centerline{\includegraphics[height=1.03\textwidth, width=1.03\textwidth]{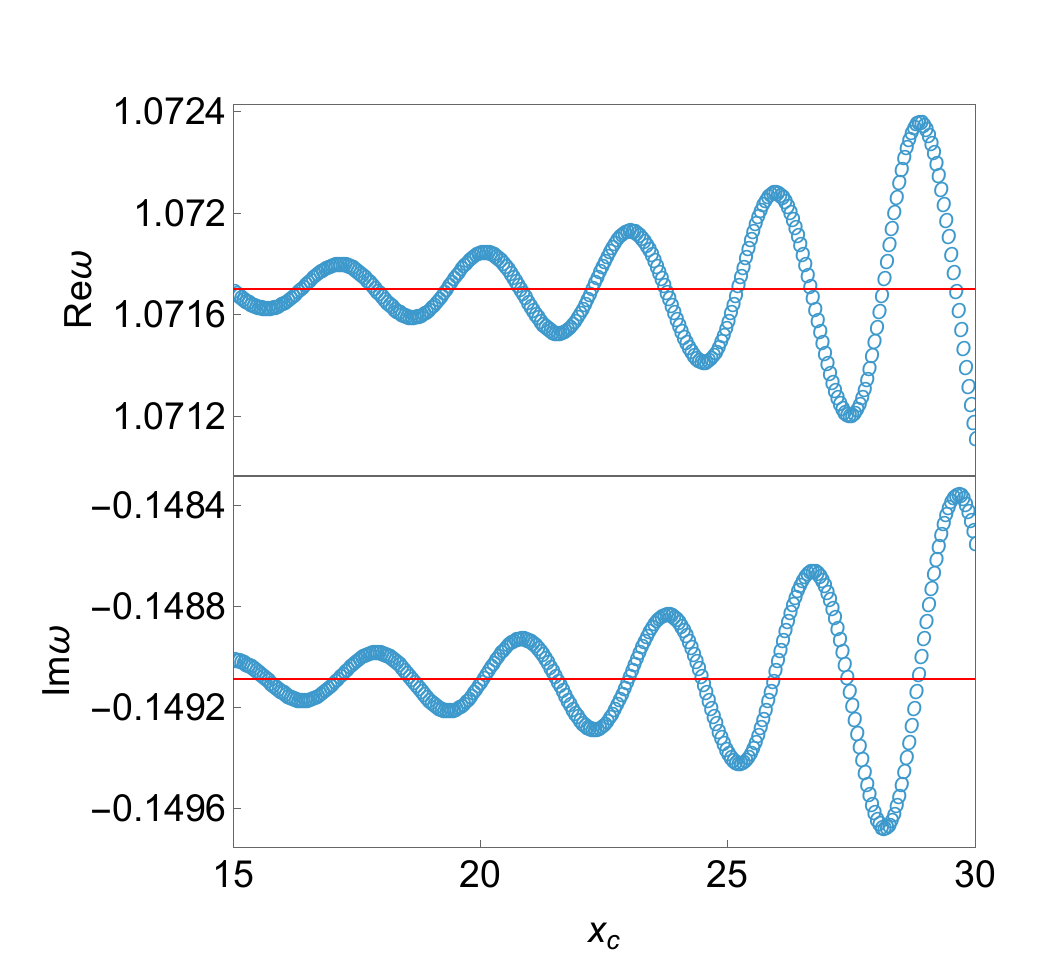}}
\end{minipage}
\end{tabular}
\renewcommand{\figurename}{Fig.}
\caption{The resulting fundamental mode as a function of the location of the metric perturbation $x_c$ in the approximate Regge-Wheeler effective potential.
As the coordinate $x_c$ varies from 15 to 30 with an increment of 0.05, the resulting fundamental mode spirals away from the original fundamental mode in the counterclockwise direction.
The fundamental modes of the perturbed metric are presented by empty blue circles, and the location of the corresponding unperturbed mode $\omega_0 = 1.0717 - i 0.14909$ is indicated by the filled red circle (left panel) and solid red lines (right panel).
The numerical calculations are carried out using the analytic wavefunctions given by Eqs.~\eqref{RWAsolf}-\eqref{RWAsoluH}, and~\eqref{RWAsolg2}, where one adopts the metric parameters $\ell=2$, $s=0$, $x_0=-0.9575$, $x_1=2.4637$, and $\epsilon=0.01$.
}
\label{RW_app_spiral} 
\end{figure}

In this section, we carry out numerical calculations that support our foregoing analytic estimations.
The results are shown in Fig.~\ref{RW_app_spiral} and Tab.~\ref{tab:omega_selected_int}.
In the calculations, one employed the full analytic expressions of the wavefunctions given in Eqs.~\eqref{RWAsolf}-\eqref{RWAsoluH}, and~\eqref{RWAsolg2}.
One plants metric perturbations to the Regge-Wheeler effective potential of a black hole with $r_h=1$ using the parameters $\ell=2$, $s=0$, $x_0=-0.9575$, $x_1=2.4637$, and $\epsilon=0.01$.

In Fig.~\ref{RW_app_spiral}, as $x_c$ increases, we observe a counterclockwise spiral away from the original, unperturbed, fundamental mode.
The instability is in agreement with the conclusion drawn in Eq.~\eqref{eqDev2}.
Quantitatively, one can perform a more detailed analysis as follows.
On the one hand, the period of the spiral is governed by the real part of the frequency. 
One observes that the frequency deviation $\delta\omega$ completes five full circles as $x_c=15$ evolves to $29.65$.
Using Eq.~\eqref{eqDev2}, equating the arguments implies $5\times 2\pi=2 (29.65-15)\times \mathrm{Re}\omega$ that gives $\mathrm{Re}\omega\sim 1.072$, where we note that $\mathscr{F}$ is not a function of $x_c$ and the denominator $9i+10\epsilon\omega^3x_c^3$ only gives a small correction ($<\pi/2$) to the phase.
The extracted value is nearly identical to $\mathrm{Re}\omega_0=1.0717$.
On the other hand, the magnitude of the deviation is related to the imaginary part of the frequency.
We can estimate the deviation in magnitude by noticing that as the mode $\omega(15)=1.07169-i0.149015$ evolves to $\omega(30)=1.07111-i0.148555$, the corresponding values of $x_c$ are $15$ and $30$.
By taking into account the $x_c$ dependence of the prefactor on the r.h.s. of Eq.~\eqref{defJ2}, this gives $\ln\left(30^3/15^3\right)\ln\left(\frac{|\omega(30)-\omega_0|}{|\omega(15)-\omega_0|}\right)=-2(30-15)\times\mathrm{Im}\omega$ and subsequently $\mathrm{Im}\omega\sim -0.16$.
This is in reasonable agreement with the imaginary part of the unperturbed fundamental mode $\mathrm{Im}\omega_0=-0.149$.
It is worth noting that if one assumes that the magnitude of the spiral follows the simple form based on Eq.~\eqref{DoubleSqureBarrier}, namely, $|\delta\omega|\propto e^{-2\mathrm{Im}\omega_0 x_c}$, the extracted imaginary part of the frequency would have been $\mathrm{Im}\omega\sim -0.079$, significantly deviating from the true value.

\begin{table}[h]
\centering
\caption{The numerical values of the quasinormal frequencies presented in Fig.~\ref{RW_app_spiral} as a function of $x_c$.}
\label{tab:omega_selected_int}
\begin{tabular}{*{9}{c}}
\hline\hline
$x_c$ & 15 & 16 & 17 & 18 & 19 & 20 & 21 & 22 \\
\hline
$\mathrm{Re}\,\omega$ & 1.07169 & 1.07164 & 1.07179 & 1.07167 & 1.07162 & 1.07184 & 1.07163 & 1.07160 \\
$\mathrm{Im}\,\omega$ & -0.149015 & -0.149145 & -0.149111 & -0.148988 & -0.149183 & -0.149103 & -0.148942 & -0.149253 \\
\hline\hline
$x_c$ & 23 & 24 & 25 & 26 & 27 & 28 & 29 & 30 \\
\hline
$\mathrm{Re}\,\omega$ & 1.07193 & 1.07155 & 1.07157 & 1.07208 & 1.07140 & 1.07155 & 1.07233 & 1.07111 \\
$\mathrm{Im}\,\omega$ & -0.149080 & -0.148867 & -0.149387 & -0.149017 & -0.148747 & -0.149644 & -0.148863 & -0.148555 \\
\hline\hline
\end{tabular}
\end{table}

\section{Concluding remarks}\label{sec5}

To summarize, we explore the instability of the fundamental mode of the Regge-Wheeler effective potential by employing an approximate analytic form for the potential.
The study aims to provide a primarily analytic account of the observed instability recently highlighted by Cheung {\it et al.}, and it is arguable that the current interpretation of this result might have oversimplified the actual physical scenario.
We pointed out that a metric perturbation planted farther away from the black hole horizon might not always be appropriately approximated by a disjoint minor barrier.
In this regard, the present study performed an analysis using explicit analytic forms of the wavefunctions.
Among others, we consider a physically relevant scenario, where the metric perturbation mimics an infinitesimal mass shell of constant mass, so that as the metric perturbation moves away, its physical impact largely remains unchanged. 
While our analysis confirms the conclusion in the literature regarding the stability of the fundamental mode, it is demonstrated that the difference between Eqs.~\eqref{DoubleSqureBarrier},~\eqref{SpiralRWApp1} and~\eqref{SpiralRWApp2} might be numerically substantial.
Specifically, for the parameters adopted for the approximate Regge-Wheeler potential, the discrepancy between an estimation and the true value of the imaginary part of the quasinormal frequency might be as large as 50\%.
Such a difference reflects the specific properties of the spacetime at spatial infinity and/or the underlying metric perturbations.
On the other hand, the spiral period is mainly governed by the real part of the quasinormal frequency, irrespective of the remaining details of the spacetime metric.
Our analytic estimations agree reasonably well with numerical calculations.

Recent progress regarding black hole spectral instability leads to discussions about its observational implications.
Despite significant deformation of the QNM spectrum, various studies~\cite{agr-qnm-instability-02, agr-qnm-instability-11, agr-qnm-instability-47, agr-qnm-instability-83, agr-qnm-echoes-48} have shown that the impact on the time-domain waveform is minimal.
From our understanding of black hole spectroscopy, many questions remain unanswered and pertinent, particularly in the context of ongoing space-borne gravitational wave detection projects.

\section*{Acknowledgements}

We gratefully acknowledge the financial support from Brazilian agencies 
Funda\c{c}\~ao de Amparo \`a Pesquisa do Estado de S\~ao Paulo (FAPESP), 
Funda\c{c}\~ao de Amparo \`a Pesquisa do Estado do Rio de Janeiro (FAPERJ), 
Conselho Nacional de Desenvolvimento Cient\'{\i}fico e Tecnol\'ogico (CNPq), 
and Coordena\c{c}\~ao de Aperfei\c{c}oamento de Pessoal de N\'ivel Superior (CAPES).
This work is supported by the National Natural Science Foundation of China (NSFC).
A part of this work was developed under the project Institutos Nacionais de Ci\^{e}ncias e Tecnologia - Física Nuclear e Aplica\c{c}\~{o}es (INCT/FNA) Proc. No. 464898/2014-5.
This research is also supported by the Center for Scientific Computing (NCC/GridUNESP) of São Paulo State University (UNESP).

\bibliographystyle{h-physrev}
\bibliography{references_qian}

\begin{thebibliography}{10}

\bibitem{agr-qnm-instability-02}
H.-P. Nollert,
\newblock Phys. Rev. {\bf D53}, 4397 (1996), arXiv:gr-qc/9602032.

\bibitem{agr-qnm-instability-03}
H.-P. Nollert and R.~H. Price,
\newblock J. Math. Phys. {\bf 40}, 980 (1999), arXiv:gr-qc/9810074.

\bibitem{agr-qnm-27}
J.~M. Aguirregabiria and C.~V. Vishveshwara,
\newblock Phys. Lett. A {\bf 210}, 251 (1996).

\bibitem{agr-qnm-30}
C.~V. Vishveshwara,
\newblock Curr. Sci. {\bf 71}, 824 (1996).

\bibitem{agr-qnm-review-01}
K.~D. Kokkotas and B.~G. Schmidt,
\newblock Living Rev. Rel. {\bf 2}, 2 (1999), arXiv:gr-qc/9909058.

\bibitem{agr-qnm-review-02}
H.-P. Nollert,
\newblock Class. Quant. Grav. {\bf 16}, R159 (1999).

\bibitem{agr-qnm-review-03}
E.~Berti, V.~Cardoso, and A.~O. Starinets,
\newblock Class. Quant. Grav. {\bf 26}, 163001 (2009), arXiv:0905.2975.

\bibitem{agr-qnm-instability-11}
R.~G. Daghigh, M.~D. Green, and J.~C. Morey,
\newblock Phys. Rev. {\bf D101}, 104009 (2020), arXiv:2002.07251.

\bibitem{agr-qnm-lq-03}
W.-L. Qian, K.~Lin, C.-Y. Shao, B.~Wang, and R.-H. Yue,
\newblock Phys. Rev. {\bf D103}, 024019 (2021), arXiv:2009.11627.

\bibitem{agr-qnm-continued-fraction-12}
H.-P. Nollert,
\newblock Phys. Rev. {\bf D47}, 5253 (1993).

\bibitem{agr-qnm-continued-fraction-23}
L.~Motl,
\newblock Adv. Theor. Math. Phys. {\bf 6}, 1135 (2003), arXiv:gr-qc/0212096.

\bibitem{agr-qnm-21}
E.~W. Leaver,
\newblock Phys. Rev. {\bf D34}, 384 (1986).

\bibitem{agr-qnm-29}
H.-P. Nollert and B.~G. Schmidt,
\newblock Phys. Rev. {\bf D45}, 2617 (1992).

\bibitem{agr-qnm-echoes-20}
H.~Liu {\em et~al.},
\newblock Phys. Rev. {\bf D104}, 044012 (2021), arXiv:2104.11912.

\bibitem{agr-strong-lensing-correlator-15}
W.-L. Qian, K.~Lin, X.-M. Kuang, B.~Wang, and R.-H. Yue,
\newblock Eur. Phys. J. C {\bf 82}, 188 (2022), arXiv:2109.02844.

\bibitem{agr-qnm-echoes-45}
S.-F. Shen {\em et~al.},
\newblock Phys. Rev. D {\bf 110}, 084022 (2024), arXiv:2408.00971.

\bibitem{agr-qnm-instability-65}
R.~G. Daghigh, G.-R. Li, W.-L. Qian, and S.~J. Randow,
\newblock Phys. Rev. D {\bf 111}, 124021 (2025), arXiv:2502.05354.

\bibitem{agr-qnm-echoes-01}
V.~Cardoso, E.~Franzin, and P.~Pani,
\newblock Phys. Rev. Lett. {\bf 116}, 171101 (2016), arXiv:1602.07309,
\newblock [Erratum: Phys. Rev. Lett.117,no.8,089902(2016)].

\bibitem{agr-qnm-echoes-review-01}
V.~Cardoso and P.~Pani,
\newblock Living Rev. Rel. {\bf 22}, 4 (2019), arXiv:1904.05363.

\bibitem{agr-qnm-instability-07}
J.~L. Jaramillo, R.~Panosso~Macedo, and L.~Al~Sheikh,
\newblock Phys. Rev. X {\bf 11}, 031003 (2021), arXiv:2004.06434.

\bibitem{agr-qnm-instability-13}
J.~L. Jaramillo, R.~Panosso~Macedo, and L.~A. Sheikh,
\newblock Phys. Rev. Lett. {\bf 128}, 211102 (2022), arXiv:2105.03451.

\bibitem{agr-BH-spectroscopy-review-04}
E.~Berti {\em et~al.},
\newblock (2025), arXiv:2505.23895.

\bibitem{agr-LIGO-01}
Virgo, LIGO Scientific, B.~P. Abbott {\em et~al.},
\newblock Phys. Rev. Lett. {\bf 116}, 061102 (2016), arXiv:1602.03837.

\bibitem{agr-LIGO-02}
Virgo, LIGO Scientific, B.~P. Abbott {\em et~al.},
\newblock Phys. Rev. Lett. {\bf 116}, 221101 (2016), arXiv:1602.03841,
\newblock [Erratum: Phys. Rev. Lett.121,no.12,129902(2018)].

\bibitem{agr-LIGO-03}
Virgo, LIGO Scientific, B.~P. Abbott {\em et~al.},
\newblock Phys. Rev. Lett. {\bf 116}, 241103 (2016), arXiv:1606.04855.

\bibitem{agr-LIGO-04}
Virgo, LIGO Scientific, B.~P. Abbott {\em et~al.},
\newblock Phys. Rev. Lett. {\bf 119}, 141101 (2017), arXiv:1709.09660.

\bibitem{agr-LISA-01}
LISA, P.~Amaro-Seoane {\em et~al.},
\newblock (2017), arXiv:1702.00786.

\bibitem{agr-TianQin-01}
TianQin, J.~Luo {\em et~al.},
\newblock Class. Quant. Grav. {\bf 33}, 035010 (2016), arXiv:1512.02076.

\bibitem{agr-TianQin-Taiji-review-01}
Y.~Gong, J.~Luo, and B.~Wang,
\newblock Nature Astron. {\bf 5}, 881 (2021), arXiv:2109.07442.

\bibitem{agr-Taiji-01}
W.-R. Hu and Y.-L. Wu,
\newblock Natl. Sci. Rev. {\bf 4}, 685 (2017).

\bibitem{agr-TianQin-05}
C.~Shi {\em et~al.},
\newblock Phys. Rev. D {\bf 100}, 044036 (2019), arXiv:1902.08922.

\bibitem{agr-bh-thermodynamics-12}
M.~Visser,
\newblock Phys. Rev. {\bf D46}, 2445 (1992), arXiv:hep-th/9203057.

\bibitem{agr-qnm-33}
P.~T. Leung, Y.~T. Liu, W.~M. Suen, C.~Y. Tam, and K.~Young,
\newblock Phys. Rev. Lett. {\bf 78}, 2894 (1997), arXiv:gr-qc/9903031.

\bibitem{agr-qnm-34}
P.~T. Leung, Y.~T. Liu, W.~M. Suen, C.~Y. Tam, and K.~Young,
\newblock Phys. Rev. {\bf D59}, 044034 (1999), arXiv:gr-qc/9903032.

\bibitem{agr-BH-spectroscopy-10}
E.~Barausse, V.~Cardoso, and P.~Pani,
\newblock Phys. Rev. {\bf D89}, 104059 (2014), arXiv:1404.7149.

\bibitem{agr-qnm-instability-08}
E.~Gasperin and J.~L. Jaramillo,
\newblock Class. Quant. Grav. {\bf 39}, 115010 (2022), arXiv:2107.12865.

\bibitem{agr-qnm-instability-14}
K.~Destounis, R.~P. Macedo, E.~Berti, V.~Cardoso, and J.~L. Jaramillo,
\newblock Phys. Rev. D {\bf 104}, 084091 (2021), arXiv:2107.09673.

\bibitem{agr-qnm-instability-15}
M.~H.-Y. Cheung, K.~Destounis, R.~P. Macedo, E.~Berti, and V.~Cardoso,
\newblock Phys. Rev. Lett. {\bf 128}, 111103 (2022), arXiv:2111.05415.

\bibitem{agr-qnm-instability-16}
E.~Berti {\em et~al.},
\newblock Phys. Rev. D {\bf 106}, 084011 (2022), arXiv:2205.08547.

\bibitem{agr-qnm-instability-18}
K.~Kyutoku, H.~Motohashi, and T.~Tanaka,
\newblock Phys. Rev. D {\bf 107}, 044012 (2023), arXiv:2206.00671.

\bibitem{agr-qnm-instability-19}
J.~L. Jaramillo,
\newblock Class. Quant. Grav. {\bf 39}, 217002 (2022), arXiv:2206.08025.

\bibitem{agr-qnm-instability-26}
H.~Yang and J.~Zhang,
\newblock Phys. Rev. D {\bf 107}, 064045 (2023), arXiv:2210.01724.

\bibitem{agr-qnm-echoes-22}
L.~Hui, D.~Kabat, and S.~S.~C. Wong,
\newblock JCAP {\bf 12}, 020 (2019), arXiv:1909.10382.

\bibitem{agr-qnm-echoes-29}
M.~Rahman and A.~Bhattacharyya,
\newblock Phys. Rev. D {\bf 104}, 044045 (2021), arXiv:2104.00074.

\bibitem{agr-qnm-echoes-30}
K.~Chakravarti, R.~Ghosh, and S.~Sarkar,
\newblock Phys. Rev. D {\bf 105}, 044046 (2022), arXiv:2112.10109.

\bibitem{agr-qnm-instability-23}
R.~A. Konoplya and A.~Zhidenko,
\newblock JHEAp {\bf 44}, 419 (2024), arXiv:2209.00679.

\bibitem{agr-qnm-instability-29}
V.~Boyanov, K.~Destounis, R.~Panosso~Macedo, V.~Cardoso, and J.~L. Jaramillo,
\newblock Phys. Rev. D {\bf 107}, 064012 (2023), arXiv:2209.12950.

\bibitem{agr-qnm-instability-32}
A.~Courty, K.~Destounis, and P.~Pani,
\newblock Phys. Rev. D {\bf 108}, 104027 (2023), arXiv:2307.11155.

\bibitem{agr-qnm-instability-33}
S.~Sarkar, M.~Rahman, and S.~Chakraborty,
\newblock Phys. Rev. D {\bf 108}, 104002 (2023), arXiv:2304.06829.

\bibitem{agr-qnm-instability-43}
V.~Boyanov, V.~Cardoso, K.~Destounis, J.~L. Jaramillo, and R.~Panosso~Macedo,
\newblock Phys. Rev. D {\bf 109}, 064068 (2024), arXiv:2312.11998.

\bibitem{agr-qnm-echoes-35}
W.-L. Qian, Q.~Pan, B.~Wang, and R.-H. Yue,
\newblock Phys. Lett. B {\bf 856}, 138874 (2024), arXiv:2402.05485.

\bibitem{spectral-instability-review-20}
S.-F. Shen {\em et~al.},
\newblock (2025), arXiv:2507.11663.

\bibitem{agr-qnm-instability-56}
Y.~Yang, Z.-F. Mai, R.-Q. Yang, L.~Shao, and E.~Berti,
\newblock Phys. Rev. D {\bf 110}, 084018 (2024), arXiv:2407.20131.

\bibitem{agr-qnm-instability-58}
A.~Ianniccari {\em et~al.},
\newblock Phys. Rev. Lett. {\bf 133}, 211401 (2024), arXiv:2407.20144.

\bibitem{agr-qnm-instability-55}
W.-L. Qian, G.-R. Li, R.~G. Daghigh, S.~Randow, and R.-H. Yue,
\newblock Phys. Rev. D {\bf 111}, 024047 (2025), arXiv:2409.17026.

\bibitem{agr-qnm-instability-59}
V.~Boyanov,
\newblock Front. in Phys. {\bf 12}, 1511757 (2024), arXiv:2410.11547.

\bibitem{agr-qnm-instability-84}
S.-F. Shen {\em et~al.},
\newblock (2025), arXiv:2509.23372.

\bibitem{agr-qnm-review-06}
B.~Wang,
\newblock Braz. J. Phys. {\bf 35}, 1029 (2005), arXiv:gr-qc/0511133.

\bibitem{agr-qnm-tail-06}
E.~S.~C. Ching, P.~T. Leung, W.~M. Suen, and K.~Young,
\newblock Phys. Rev. {\bf D52}, 2118 (1995), arXiv:gr-qc/9507035.

\bibitem{agr-qnm-instability-47}
V.~Cardoso, S.~Kastha, and R.~Panosso~Macedo,
\newblock Phys. Rev. D {\bf 110}, 024016 (2024), arXiv:2404.01374.

\bibitem{agr-qnm-instability-83}
L.-B. Wu, L.~Xie, Y.-S. Zhou, Z.-K. Guo, and R.-G. Cai,
\newblock (2025), arXiv:2509.20947.

\bibitem{agr-qnm-echoes-48}
Z.-H. Yang, L.-B. Wu, X.-M. Kuang, and W.-L. Qian,
\newblock (2025), arXiv:2510.02033.

\end{thebibliography}

\end{document}